\documentclass[12pt,a4paper]{article}
\usepackage{amsmath}
\begin{document}

\title{\textbf{B\"{a}cklund transformation}\\%
\textbf{and special solutions for}\\%
\textbf{Drinfeld--Sokolov--Satsuma--Hirota}\\%
\textbf{system of coupled equations}}
\author{\textsc{Ay\c{s}e Karasu (Kalkanli)}\thanks{E-mail:
\textit{akarasu@metu.edu.tr}} \ and \textsc{S Yu Sakovich}\thanks{Permanent
address: \textit{Institute of Physics, National Academy of Sciences, 220072
Minsk, Belarus}. E-mail: \textit{saks@pisem.net}}\medskip\\%
\textit{Department of Physics, Middle East Technical University,}\\%
\textit{06531 Ankara, Turkey}}
\date{}
\maketitle

\begin{abstract}
Using the Weiss method of truncated singular expansions, we construct an
explicit B\"{a}cklund transformation of the Drinfeld--Sokolov--Satsuma--Hirota
system into itself. Then we find all the special solutions generated by this
transformation from the trivial zero solution of this system.\medskip

Short title: \textit{B\"{a}cklund transformation and special solutions}

PACS numbers: \textit{02.30.Ik, 02.30.Jr\bigskip}

\end{abstract}

The system of two coupled nonlinear evolution equations%
\begin{equation}
u_{t}+u_{xxx}-6uu_{x}-6v_{x}=0\qquad v_{t}-2v_{xxx}+6uv_{x}=0 \label{dssh}%
\end{equation}
was introduced, independently, by Drinfeld and Sokolov \cite{DS}, and by
Satsuma and Hirota \cite{SH}. In \cite{DS}, the system (\ref{dssh}) was given
as one of numerous examples of nonlinear equations possessing Lax pairs of a
special form. In \cite{SH}, the system (\ref{dssh}) was found as a special
case of the four-reduction of the KP hierarchy, and its explicit one-soliton
solution was constructed. Recently, G\"{u}rses and Karasu \cite{GK} found a
recursion operator and a bi-Hamiltonian structure for (\ref{dssh}), which
provide the system (\ref{dssh}) with an infinite algebra of generalized
symmetries and an infinite set of conservation laws.

In this paper, we construct, using the Weiss method of truncated singular
expansions \cite{Wei}, an explicit B\"{a}cklund transformation of the
Drinfeld--Sokolov--Satsuma--Hirota (DSSH) system (\ref{dssh}) into itself, and
then find all the special solutions generated by this transformation from the
trivial zero solution of this system. In relation with our results, we should
mention the recent paper of Tian and Gao \cite{TG}, who did not find a
B\"{a}cklund transformation of (\ref{dssh}) into itself, contrary to their
claim, and produced a rational special solution of the DSSH system with
erroneously many arbitrary parameters.

First of all, we find it useful to rewrite the DSSH system (\ref{dssh}) in the
form of the single sixth-order equation%
\begin{equation}
w_{tt}-w_{xxxt}-2w_{xxxxxx}+18w_{x}w_{xxxx}+36w_{xx}w_{xxx}-36w_{x}^{2}%
w_{xx}=0 \label{eq}%
\end{equation}
which is related to (\ref{dssh}) by the Miura-type transformation%
\begin{equation}
u=w_{x}\qquad v=\frac{1}{6}\left(  w_{t}+w_{xxx}-3w_{x}^{2}\right)  .
\label{uv}%
\end{equation}
This representation (\ref{eq}) of the original system (\ref{dssh}) is very
convenient, because the B\"{a}cklund transformation of (\ref{dssh}) would
contain complicated radicals, whereas the B\"{a}cklund transformation of
(\ref{eq}) turns out to involve rational expressions only.

It is easy to verify that the equation (\ref{eq}) possesses the Painlev\'{e}
property in the formulation for partial differential equations \cite{WTC}.
Substituting the singular expansion $w=w_{0}\phi^{\alpha}+\cdots+w_{n}%
\phi^{n+\alpha}+\cdots$ into (\ref{eq}), we find the following two branches,
i.e.\ admissible choices of $\alpha$ and $w_{0}$ with corresponding positions
$n$ of resonances: (i) $\alpha=-1$, $w_{0}=-2\phi_{x}$, $n=-1,1,3,4,6,8$, and
(ii) $\alpha=-1$, $w_{0}=-10\phi_{x}$, $n=-5,-1,1,6,8,12$. Then, checking the
consistency of the recursion relations for $w_{n}$ at the resonances of both
branches, we find that no logarithmic terms should be introduced into the
singular expansions of solutions.

According to the positions of resonances, the branch (i) is generic, but the
branch (ii) is not. Since the B\"{a}cklund transformation sought should be
applicable to a generic solution of the equation (\ref{eq}), we have to
consider the truncated singular expansion in the branch (i). The use of the
new expansion function $\chi$, $\chi=\left(  \phi^{-1}\phi_{x}-\frac{1}{2}%
\phi_{x}^{-1}\phi_{xx}\right)  ^{-1}$, proposed by Conte \cite{Con},
simplifies the computations very considerably. Following \cite{Con}, we
substitute the truncated expansion $w=-2\chi^{-1}+a\left(  x,t\right)  $ into
the equation (\ref{eq}), use the identities $\chi_{x}=1+\frac{1}{2}S\chi^{2}$,
$\chi_{t}=-C+C_{x}\chi-\frac{1}{2}\left(  C_{xx}+CS\right)  \chi^{2}$ and
$S_{t}+C_{xxx}+2SC_{x}+CS_{x}=0$, where $S=\phi_{x}^{-1}\phi_{xxx}-\frac{3}%
{2}\phi_{x}^{-2}\phi_{xx}^{2}$ and $C=-\phi_{x}^{-1}\phi_{t}$, collect terms
with equal degrees of $\chi$, and thus obtain a complicated system of
equations, which turns out to be compatible and equivalent to the following
normal system of two equations for $\phi$ and $a$ of total order six:%
\begin{equation}
a_{x}+\frac{1}{6}C+\frac{1}{3}S=0\qquad a_{t}+a_{xxx}-\frac{3}{2}a_{x}%
^{2}+\frac{3}{2}a_{x}C+\frac{3}{8}C^{2}=k \label{aux}%
\end{equation}
where the parameter $k$ appeared as a constant of integration.

The truncated singular expansion%
\begin{equation}
w=-2\frac{\phi_{x}}{\phi}+\frac{\phi_{xx}}{\phi_{x}}+a \label{w}%
\end{equation}
represents a Miura-type transformation between the system (\ref{aux}) and the
equation (\ref{eq}). According to the Weiss method \cite{Wei}, the function
$z\left(  x,t\right)  $ determined by%
\begin{equation}
z=\frac{\phi_{xx}}{\phi_{x}}+a \label{z}%
\end{equation}
is also a solution of (\ref{eq}). Therefore (\ref{z}) represents one more
Miura-type transformation between (\ref{aux}) and (\ref{eq}). These two
Miura-type transformations, (\ref{w}) and (\ref{z}), together with the
auxiliary system (\ref{aux}), determine an implicit B\"{a}cklund
transformation between the `new' solution $w$ and the `old' solution $z$ of
the DSSH system written in the form (\ref{eq}) (certainly, these terms `new'
and `old' can be used in the opposite order).

Now, eliminating $\phi$ and $a$ from the equations (\ref{aux}), (\ref{w}) and
(\ref{z}), we obtain the following explicit B\"{a}cklund transformation of the
equation (\ref{eq}) into itself:%
\begin{equation}
p_{t}+\left(  4p_{xx}-3\frac{p_{x}^{2}}{p}-3pq_{x}+\frac{1}{4}p^{3}\right)
_{x}=0 \label{bt1}%
\end{equation}%
\begin{gather}
q_{t}-\frac{3}{2}\frac{p_{xt}}{p}+\frac{3}{4}\frac{p_{x}p_{t}}{p^{2}}-\frac
{1}{2}q_{xxx}-\frac{3}{2}\frac{q_{x}p_{xx}}{p}+\frac{3}{4}\frac{p_{x}q_{xx}%
}{p}+\frac{9}{8}pp_{xx}\nonumber\\
+\frac{3}{4}\frac{p_{x}^{2}q_{x}}{p^{2}}-\frac{9}{16}p_{x}^{2}+\frac{3}%
{2}q_{x}^{2}-\frac{3}{4}p^{2}q_{x}+\frac{3}{64}p^{4}=2k \label{bt2}%
\end{gather}
where $p=w-z$ and $q=w+z$. The equations (\ref{bt1}) and (\ref{bt2})
constitute a B\"{a}cklund transformation in the sense of the definition given
in \cite{AS}: if we eliminate $z$ from (\ref{bt1}) and (\ref{bt2}), we obtain
exactly the equation (\ref{eq}) for $w$; and if we eliminate $w$ from
(\ref{bt1}) and (\ref{bt2}), we obtain (\ref{eq}) for $z$. We think, however,
that it is impossible to prove this by hand-made computations: we did it by
means of the \textit{Mathematica} system \cite{Wol}. As far as we know, the
equations (\ref{bt1}) and (\ref{bt2}) are the most complicated explicit
B\"{a}cklund transformation in the literature.

Let us find all the `new' solutions $w$ of the equation (\ref{eq}) generated
by the obtained B\"{a}cklund transformation from the trivial `old' solution
$z=0$. Setting $z=0$ in (\ref{bt1}) and (\ref{bt2}), we get an over-determined
system of two equations for $w$, which can be solved exactly, using the new
dependent variable $\psi$: $w=-2\phi^{-1}\phi_{x}$, $\phi_{x}=\psi^{2}$. This
gives us the following five types of explicit special solutions of the DSSH
system written in the form (\ref{eq}):%
\begin{equation}
w=\frac{-2}{x} \label{s1}%
\end{equation}%
\begin{equation}
w=\frac{-6x^{2}}{x^{3}+12t} \label{s2}%
\end{equation}%
\begin{equation}
w=\frac{-30\left(  x^{2}+\sigma\right)  ^{2}}{3x^{5}+10\sigma x^{3}%
+15\sigma^{2}x-240\sigma t+\tau} \label{s3}%
\end{equation}%
\begin{gather}
w=-210\left(  x^{3}+\sigma x-24t\right)  ^{2}\left(  15x^{7}+42\sigma
x^{5}-1260x^{4}t\right. \nonumber\\
\left.  +35\sigma^{2}x^{3}-2520\sigma x^{2}t+60480xt^{2}+420\sigma^{2}%
t+\tau\right)  ^{-1} \label{s4}%
\end{gather}%
\begin{gather}
w=-4\kappa\left(  c_{1}\mathrm{e}^{a}+c_{2}\mathrm{e}^{-a}+c_{3}\mathrm{e}%
^{b}+c_{4}\mathrm{e}^{-b}\right)  ^{2}\nonumber\\
\times\left[  c_{1}^{2}\mathrm{e}^{2a}-c_{2}^{2}\mathrm{e}^{-2a}%
-\mathrm{i}c_{3}^{2}\mathrm{e}^{2b}+\mathrm{i}c_{4}^{2}\mathrm{e}^{-2b}\right.
\nonumber\\
+2\left(  1-\mathrm{i}\right)  c_{1}c_{3}\mathrm{e}^{a+b}+2\left(
1+\mathrm{i}\right)  c_{1}c_{4}\mathrm{e}^{a-b}\nonumber\\
-2\left(  1+\mathrm{i}\right)  c_{2}c_{3}\mathrm{e}^{-a+b}-2\left(
1-\mathrm{i}\right)  c_{2}c_{4}\mathrm{e}^{-a-b}\nonumber\\
\left.  +4\left(  c_{1}c_{2}+c_{3}c_{4}\right)  \kappa x-48\left(  c_{1}%
c_{2}-c_{3}c_{4}\right)  \kappa^{3}t+c_{5}\right]  ^{-1} \label{s5}%
\end{gather}
where some inessential parameters were eliminated using arbitrary shifts
$x\rightarrow x+x_{0}$ and $t\rightarrow t+t_{0}$; $\sigma,\tau,c_{1}%
,c_{2},c_{3},c_{4},c_{5}$ and $\kappa$ are arbitrary constants; $a=\kappa
x-4\kappa^{3}t$ and $b=\mathrm{i}\kappa x+4\mathrm{i}\kappa^{3}t$; $\kappa$ is
related to $k$ as $k=6\kappa^{4}$; and the solutions (\ref{s1})--(\ref{s4})
correspond to the case $k=0$.

With respect to the DSSH system (\ref{dssh}), the solutions (\ref{s1}) and
(\ref{s2}) are trivial in the sense that, according to (\ref{uv}), $v=0$ for
them; actually, they are rational solutions of the potential KdV equation
$w_{t}+w_{xxx}-3w_{x}^{2}=0$. But the rational solutions (\ref{s3}) and
(\ref{s4}) are not trivial in this sense: $v\neq0$ for them. As for the
solution (\ref{s5}), $v=0$ only if $c_{1}=c_{2}=0$ or $c_{3}=c_{4}=0$.
Consequently, the obtained B\"{a}cklund transformation can produce nontrivial
solutions for the coupled equations (\ref{dssh}).

We should notice, however, that there are no solitary wave solutions among the
solutions (\ref{s1})--(\ref{s5}). Indeed, the only solution of the form
$w=f\left(  x-ct\right)  $ ($c=\mathrm{constant}$), generated by
(\ref{bt1})--(\ref{bt2}) from $z=0$, is the solution (\ref{s1}), for which
$c=0$. For this reason, we suspect (\ref{bt1})--(\ref{bt2}) of being not the
simplest (elementary) B\"{a}cklund transformation of the equation (\ref{eq})
into itself: it might be a product of two elementary B\"{a}cklund
transformations; such a phenomenon was observed in \cite{Sak}, where two
different B\"{a}cklund transformations of the Calogero equation were found and
studied. This point requires further investigation.\bigskip

The authors are grateful to the Scientific and Technical Research Council of
Turkey (T\"{U}B\.{I}TAK) for support.

\end{document}